\documentclass[aps,prb,twocolumn,superscriptaddress,showpacs]{revtex4}

\usepackage{amsmath,bm}
\usepackage{graphicx}

\begin{document}

\title{NRG for the bosonic single-impurity Anderson model: Dynamics}

\author{Hyun-Jung Lee}
\affiliation{Asia Pacific Center for Theoretical Physics, POSTECH, Pohang, \it Korea}
\author{Krzysztof Byczuk}
\affiliation{Institute of Theoretical Physics, University of Warsaw, ul. Ho\.za 69, 00-681 Warszawa,
{\it Poland}}
\author{Ralf Bulla}
\affiliation{Institut f\"ur Theoretische Physik, Universit\"at zu K\"oln,
  K\"oln, \it Germany}
\date{Draft: \today }

\begin{abstract} 
 The bosonic single-impurity Anderson model (B-SIAM) is studied to understand the local
dynamics of an atomic quantum dot (AQD) coupled to a Bose-Einstein condensation (BEC) state, which can be implemented to probe the entanglement and the decoherence of a macroscopic
condensate. Our recent approach of the numerical renormalization group (NRG) calculation for
the B-SIAM revealed a zero-temperature phase diagram, where a $Mott$ phase
with local depletion of normal particles is separated from a
$BEC$ phase with enhanced density of the condensate. As an extension of the
previous work, we present the calculations of the local dynamical quantities of
the B-SIAM which reinforce our understanding of the physics in the Mott and
the BEC phases.  

\end{abstract}
\pacs{PACS numbers: }

\maketitle
\section{Introduction}
\label{sec:intro} 
The observation of the Bose-Einstein condensation (BEC) in ultracold, atomic
gases has greatly stimulated research on the properties of this fascinating
quantum state of matter.~\cite{Anglin2002} 

A particular interest lies in the controlled manipulation of the coherence and the
entanglement of the BEC
state,~\cite{Simon2002,Heaney2008,Heaney2009A,Heaney2009B} which provide the basis of applications such as quantum computing and quantum
communications.~\cite{Nielsen2000} As an example, a new scheme of performing quantum dense coding~\cite{Bennett1992} and
teleportation~\cite{Bennett1993} was proposed using the spatial-mode entanglement of a single massive
boson coupled to a BEC reservoir.~\cite{Heaney2009A,Heaney2009B} Here one
considers a system of two coupled tightly confined potentials, each of which forms
one of the spatial modes, $A$ and $B$. In order for the full dense coding
protocol to work, $A$ and $B$ have to share a common reference frame with
which they can exchange particles. A BEC consisting of an indefinite number of
particles fulfills this role~\cite{Dowling2006} and the coherent control of
the BEC state is essential to let a signal between $A$ and $B$ be
phase-locked.

On the other hand, there have been extensive studies of decoherence, a process
of loosing quantum superpositions due to entanglement between a microscopic system and its
environment.~\cite{Zurek1991} The decoherence mechanism is crucial to
understand the transition between quantum and classical
systems~\cite{Leggett1985,Leggett2002} in a sense that quantum superposition
between distinct states of macroscopic systems is suppressed by the decoherence
process. However the environmental
effects~\cite{Zurek1991,Leggett1983} make it difficult to probe of the
decoherence of macroscopic system directly. As an alternative way, one can use the coupling of a
microscopic system, such as an atomic quantum dot, to a mesoscopic or macroscopic system to probe the
decoherence of the latter. For instance, there have been several proposals on
the single-atom-aided probe of the decoherence of a BEC.~\cite{Ng2008,Bruderer2006,Recati2005} 

In the theoretical schemes above, a BEC state is represented as the Bose-field operator, $\hat{\Psi}_{c}({\bf x})\sim
\hat{\rho}({\bf x})^{1/2}e^{-i\hat{\phi}({\bf x})}$ with the density
$\hat{\rho}({\bf x})$ and the phase $\hat{\phi}({\bf x})$ of the condensate, of which the only
available excitations at low energies are phonons with linear
dispersion. However the excitations of a BEC state is phononlike only for wavelengths
larger than the healing length $\xi$, where the healing length $\xi$ is defined as the distance over which the
condensate wave function grows
from zero to the bulk value. In general, the strong
collisional interaction in the atomic quantum dot (AQD) can locally break a BEC state to bring up
the excitations of normal particles inside of the dot. The bosonic
single-impurity Anderson model (B-SIAM)~\cite{Lee2007}  is proposed to describe the normal
excitations in the AQD as well as the condensate part.

Another motivation for studing the B-SIAM comes from a treatment of the
Bose-Hubbard model within the dynamical mean-field theory
(DMFT).~\cite{Metzner1989,Georges1996} The DMFT is an exact theory in infinite
spatial dimensions~\cite{Metzner1989} but, as an approximation for finite
dimensional system, it was successful to provide comprehensive understanding
about strongly correlated
fermion systems. Recently, a new framework of the bosonic DMFT (B-DMFT) was
proposed by Byczuk and Vollhardt~\cite{Byczuk2008} in order to extend the idea
of the DMFT to correlated lattice bosons and mixtures of bosons and fermions on a lattice.\cite{Byczuk2009}
 In contrast to the fermionic DMFT (F-DMFT), the
lattice model for bosons is mapped into a single-impurity problem with two
species of bath spectra, those from the condensate bosons and those from the
normal bosons, each of which should be self-consistently determined.
The resulting effective bosonic impurity model has been solved by
exact diagonalization (ED) method.~\cite{Hu2009,Hubener2009} Results have been
presented for various phases at finite temperatures and compared to other
theories and the experiments.

The structure of the effective impurity model in the B-DMFT is reduced to the B-SIAM
in the absence of the bath spectrum from the condensate bosons, which is the
case in the Mott insulating (MI) phase. On the basis of the current work, it
will be possible to perform NRG calculations for the B-SIAM
with a self-consistently determined bath and investigate transitions between
the superfluid (SF) and
Mott insulating (MI) phases from the side of MI phase.     

The most part in this paper is devoted to discuss the impurity quantum
phase transitions of the B-SIAM in terms of the local spectral density. In
addition, we present in detail the implementation of the bosonic NRG for the
B-SIAM to discuss various strategies to set up the iteration scheme for the bosonic NRG. 

The paper is
organized as follows: In section~\ref{sec:Model}, we introduce the Hamiltonian of
the B-SIAM, explaining the differences to the spin-boson model that has been widely used
to study the AQD coupled to a superfluid Bose-Einstein
condensate.~\cite{Recati2005} In section \ref{sec:Method}, the formulation of
the NRG for the B-SIAM is described in detail. In section~\ref{sec:phases}, we discuss the
impurity quantum phase transition of the B-SIAM and explain how the Mott and
the BEC phases are discerned in the NRG method. In section~\ref{sec:localdynamics}, we
turn to the calculation of the local spectral density to discuss the different
dynamical properties in Mott and BEC
phases. Secion~\ref{sec:conclusion} is a conclusion. We put some technical
details in appendices.
 
\section{Model Hamiltonian}
\label{sec:Model}

The spin-boson model has been widely used
for investigating the
physical properties of an atomic quantum dot (AQD) coupled to a bosonic
reservoir.~\cite{Heaney2009A,Heaney2009B,Ng2008,Bruderer2006,Recati2005}  In
Sec.~\ref{subsec:AQD}, we summarize the work by Recati {\it et
  al.}~\cite{Recati2005}, where the particle-exchange between the AQD and the
BEC reservoir has been discussed in terms of the
spin-boson model. The B-SIAM model is proposed
to relax the theoretical constrains in the spin-boson model and describe the
density fluctuation of the coherent state originated from the collisional
interaction in the AQD. In Sec.~\ref{subsec:Model}, we discuss the basic
set-up of the B-SIAM and make a comparision with the spin-boson model.

 \begin{figure}
\begin{center}
\vspace{0.cm}
  \includegraphics[angle=0,width=0.38\textwidth]{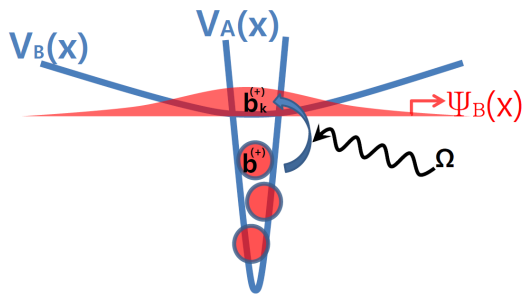}
\end{center}
\vspace{0.2cm}
\caption{Schematic setup of an atomic quantum dot coupled to a superfluid
  atomic reservoir. The
  non-interacting bose particles, denoted by the operator $b_k^{(\dagger)}$,
  are confined in a shallow trap $V_B({\bf x})$ and, at zero temperature, condense at the lowest
  vibrational mode to form a BEC state $\Psi_B({\bf x})$ (depicted as a broad wave packet). The creational operator $b^{\dagger}$
  add an atom, (depicted by balls) in the tightly confining potential $V_A({\bf x})$, where macroscopic condensation is prevented due to  the
  on-site repulsion $U$. The atoms in $V_B({\bf x})$ and $V_A({\bf x})$ are
  coupled via a Raman transition with effective Rabi frequency $\Omega$. The
  confining potential $V_A({\bf x})$ and $V_B({\bf x})$ are in all three
  directions with spherical symmetry.}
\label{opttrap1}
\end{figure}

\subsection{Atomic quantum dot coupled to a superfluid Bose-Einstein
  condensate}
\label{subsec:AQD}

The particle-exchange mechanism between an AQD and a BEC reservoir was initially proposed by
Recati {\it et al.}~\cite{Recati2005} The Hamiltonian is written as
\begin{equation}
H=H_B+H_A+H_{AB},
\label{H_org}
\end{equation}
where $H_B$ and $H_A$ correpond the energy of the BEC reservoir and the AQD,
respectively.
 The third term $H_{AB}$ describes the
Raman coupling between the AQD and the BEC.  

The first term $H_B$ describes the dynamics of the BEC reservoir. Here the reservoir atoms are assumed to form a
coherent matter wave, held in a shallow trapping potential $V_B({\bf x})$ as
illustrated in Fig.~\ref{opttrap1}. The
wave-function of the coherent state is
represented as the Bose-field operator, $\hat{\Psi}_{B}({\bf x})\sim
\hat{\rho}({\bf x})^{1/2}e^{-i\hat{\phi}({\bf x})}$, with the density
$\hat{\rho}({\bf x})$ and the phase $\hat{\phi}({\bf x})$ of the condensate.

At very low temperature, the coherent matter wave is regarded as superfluid Bose liquid with an equilibrium liquid density $\rho_B$, of which the only
available excitations are then phonons of low energy $\omega_{\bf q}=v_s|{\bf q}|$ with sound velocity $v_s$.
In this case, the dynamics of the coherent matter wave is described by a
hydrodynamic Hamiltonian,~\cite{Lifshitz1980}
\begin{equation}
H_B=\frac{1}{2}\int d{\bf x} \left(\frac{\hbar^2}{m}\rho_B|\nabla{\hat{\phi}}({\bf
  x})|^2 +\frac{mv_s^2}{\rho_B}\hat{\Pi}^2(x) \right)
\label{H_B}
\end{equation}
where $\rho_B$ is the density of the superfluid fraction and $\hat{\Pi}({\bf x})$ is
the density fluctuation operator $\hat{\Pi}({\bf x})=\hat{\rho}_B({\bf
  x})-\rho_B$, a canonical conjugate of of the superfluid phase
$\hat{\phi}({\bf x})$. The quadratic Hamiltonian in Eq.~(\ref{H_B}) can be written
in terms of standard phonon operators $b_{\bf q}$ as
\begin{equation}
H_B=\hbar v_s \sum_{\bf q}|{\bf q}| b_{\bf q}^\dagger b_{\bf q}
\end{equation}
 via the following transformation,~\cite{Recati2005}
\begin{eqnarray}
\hat{\phi}({\bf x})&=&i \sum_{\bf q} |\frac{m v_s}{2\hbar {\bf
    q}V\rho_B}|^{1/2}e^{i\bf q \cdot x}(b_{\bf q}-b_{\bf -q}^\dagger),
\nonumber \\
\hat{\Pi}({\bf x})&=&\sum_{\bf q} |\frac{\hbar\rho_B \bf q}{2 v_s V m}|^{1/2}e^{i\bf q \cdot x}(b_{\bf q}+b_{\bf -q}^\dagger).
\end{eqnarray} 
Here $V$ is the sample volume.

The second term $H_A$ corresponds to the on-site energy of the AQD.
The AQD is formed by trapping atoms in an additional tightly confining potential $V_A({\bf
x})$ as shown in Fig.~\ref{opttrap1}. Here one only considers the lowest
vibrational mode in the AQD assuming that other higher
vibrational modes are off resonant due to large detuning. The
collisional interaction of the atoms trapped in the tightly confining potential
$V_A({\bf x})$ is described by a coupling parameter $g_{AA}=4\pi
a_{AA}\hbar^2/m$ with scattering lengths $a_{AA}$ and atomic mass
$m$. The strength of the collisional interaction between the
internal states in the AQD and the coherent state in the BEC reservoir is given
as $g_{AB}=4\pi a_{AB}\hbar^2/m$. One assumes that atoms in the reservoir are
non-interacting. Thus the on-site interaction at the AQD-site is
given as,
\begin{equation}
H_A=\left[-\hbar \delta +{g_{AB}\int d{\bf x} |\psi_b({\bf x})|^2 \hat{\rho}_B({\bf
      x})}\right] \hat{b}^\dagger\hat{b}+ \frac{U_{AA}}{2}\hat{b}^\dagger\hat{b}^\dagger\hat{b}\hat{b},
\end{equation}
where $\delta$ is the detuning parameter and $\psi_A({\bf x})$ is the wave
function of the lowest vibrational mode of the AQD. The on-site repulsion in
the AQD is
given by the parameter $U_{AA}\sim g_{AA}/l_A^3$ with $l_A$ the size of the
ground state wave function $\psi_A({\bf x})$.

The last term $H_{AB}$ in Eq.~(\ref{H_org}) is the laser induced hybridization between
particles in the AQD and the BEC reservoir with effective Rabi frequency
$\Omega$:
\begin{equation}
  H_{AB}=\hbar \Omega\int {d\bf x} (\hat{\Psi}_B({\bf x})\hat{\psi}_A^\dagger({\bf x})
  + h.c.).
\label{raman}
\end{equation}
The operator $\hat{\psi}_A({\bf x})$ creates an
atom in the AQD and the operator $\hat{\Psi}_B({\bf x})$ is the annihilation operator for a reservoir atom
at the position ${\bf x}$. 

The Hamiltonian in Eq.~(\ref{H_org}) can be reduced to the spin-boson
Hamiltonian~\cite{Leggett1987} under the following conditions. 
First, one considers the collisional blockade limit of large on-site interaction $U_{AA}$,
where only states with occupation $n_A=0$ and $1$ in the AQD participate in
the dynamics. In this case the internal state of the AQD is described by a
pseudospin-1/2, with the spin-up or spin-down state corresponding to
occupation by a single or by no atom in the AQD. Using the Pauli matrix
notation, the AQD occupation operator $\hat{b}^\dagger\hat{b}$ is then
replaced by $(1+\sigma_z)/2$ while $\hat{b}^\dagger\rightarrow \sigma_+$.  

For the BEC state, one assumes that the
number of condensate atoms inside the confinement (or the AQD) is much larger than $1$, $n_B=\rho_B
l_A^3 \gg 1$; i.e., the size of the spatial confinement $l_A$ is much larger
than the average interparticle spacing in the BEC reservoir. Taking
the long wave length appriximation, $|{\bf q}|l_A\ll 1$, the phonon field
operators in $H_A$ and $H_{AB}$ are replaced by their values at ${\bf
  x}=0$. Further, neglecting the density fluctuations in the Raman coupling
in Eq.~(\ref{raman}), the Hamiltonian $H_A$ and $H_{AB}$ can be simplified to 
\begin{eqnarray}
H_A+H_{AB}&=&\left(-\frac{\hbar\delta}{2}+\frac{g_{ab}}{2}\hat{\Pi}(0)\right)\sigma_z\nonumber
\\
&&+\frac{\hbar\Delta}{2}\left(\sigma_+
e^{-i\hat{\phi}(0)}+h.c.\right).
\end{eqnarray}
Here $\Delta\sim \Omega n_B^{1/2}$ is an effective Rabi frequency.
 Eventually, after a unitary transformation $H=S^{-1}(H_A+H_B+H_{AB})S$ with
 $S=\exp\{-\sigma_z i \hat{\phi}(0)\}$, the particle-exchange mechanism between a confined boson in AQD and a
boson in the BEC reservoir can be described
by the spin-boson Hamiltonian,
\begin{eqnarray}
H&=&-\frac{\hbar \Delta}{2} \sigma_x +\sum_{\bf q} \hbar \omega_{\bf q} b_{\bf
  q}^\dagger b_{\bf q}\nonumber \\
&&+ \left[ -\delta + \sum_{\bf q} \lambda_{\bf q} (b_{\bf
    q}+b_{\bf q}^\dagger)\right]\frac{\hbar\sigma_z}{2}.
\label{sbmodel}
\end{eqnarray} 

Here the collisional interactions and
those arising from the coupling of the Rabi term to the condensate phase add
coherently in the amplitudes of the phonon coupling
\begin{equation}
 \lambda_{\bf
   q}=|\frac{m\hbar{\bf q}v_s^3}{2V\rho_B}|^{1/2}\left(\frac{g_{AB}\rho_B}{m v_s^2}-1\right).
\end{equation}

\subsection{The bosonic single-impurity Anderson model}
\label{subsec:Model} 

The Hamiltonian of the B-SIAM~\cite{Lee2007} is written as
\begin{eqnarray}
     H &=& \varepsilon {b}^\dagger {b}+\frac{U}{2}{b}^\dagger {b} ({b}^\dagger
     {b}-1)+\sum_{k}\varepsilon_k {b}^\dagger_{k}{b}_{k}\nonumber \\
     &+&\Omega\sum_{k}({b}^\dagger {b}_{k}+{b}^\dagger_{k}{b}),
\label{bsiam-H}
\end{eqnarray}
where ${b}$ and ${b}^{\dagger}$ are annihilation and creation operators obeying
bosonic canonical commutation relations and correspond to bosons within a
tight trapped potential $V_A({\bf x})$, i.e. an AQD.  The operators ${b}_k$ and ${b}_k^{\dagger}$ are  annihilation and
creation operators corresponding to non-interacting bosons confined in a shallow
potential $V_B({\bf x})$. Fig.~\ref{opttrap1} shows the schematic setup.

The energy of the AQD is given by $\varepsilon$ and $U$ is the local repulsion energy 
when two or more bosons occupy the dot system. The two parameters depend
on the strength of the collisional interaction $g_{\alpha\beta}=4\pi a_{\alpha\beta} \hbar^2/m$ with
scattering length $a_{\alpha\beta}$ ($\alpha,\beta=A$ or $B$) and the Raman detuning  $\delta$ as
discussed in Sec.~\ref{subsec:AQD}.

The third term in Eq.~(\ref{bsiam-H}) is the kinetic energy of non-interacting
bosons confined in the shallow potential $V_B({\bf x})$. Here we emphasize
that the origin of the bosonic excitations in the
B-SIAM is no more restricted to the phonons of the condensate wave function in
the lowest vibrational mode in $V_B({\bf x})$. Instead, it involves the excited particles to arbitrary higher vibrational modes in the shallow trapping potential $V_B({\bf
  x})$. The number of the vibrational modes in
  $V_B({\bf x})$ becomes infinite as the curvature of the trapping potential
  approaches to zero. In this case, the shallow trapping potential $V_B({\bf
  x})$ containing free bosons is regarded as an infinite size of a bosonic bath, of which the lowest vibrational mode
  has zero-energy. 

The last term in Eq.~(\ref{bsiam-H}) is the laser induced hybridization between
particles in the AQD and the bosonic bath with effective Rabi frequency
$\Omega$. In analogy to the fermionic SIAM the dispersion relation is determined by a hybridization function whose imaginary part, so called bath spectral function, is given by 
\begin{eqnarray}
  J(\omega)&=&\pi \Omega^2 \sum_{k}  \delta(\omega-\varepsilon_k).
\end{eqnarray}
In the following we are interested in systems with gapless bath spectral functions and in low-energy properties. Therefore, we use a model spectral function in the form 
\begin{eqnarray}
%  J_{se}(\omega)&=&\frac{V^2}{2}\sqrt{\omega(4t-\omega)}, \ 0<\omega<4t,
 % \label{bath-spctdst0}\\
  J(\omega)&=&\pi\Omega^2 (1+s)\ \omega_c^{-1-s}\omega^s \Theta(\omega_c-\omega),
  \label{bath-spctdst1}
\end{eqnarray}
where $\Theta(x)$ is a step-like theta function with a cut-off parameter 
$\omega_c$, which yields the total spectral weight $\int_0^{\omega_c}
J(\omega) d\omega=\pi \Omega^2$. Note that the choice $\omega_c=1$ sets
the energy units hereafter. The exponent $s$ characterizes how the bath
spectral functions behave in the low-energy regime.

Contrary to the spin-boson model in Ref.~(\onlinecite{Recati2005}), the B-SIAM
can consider a case where the strong Raman coupling $\Omega$ induces large density
fluctuations around the AQD-site. The Raman coupling
term in Eq.~(\ref{bsiam-H}) imposes the spatial displacement to the harmonic
oscillators in the bath, which, in
consequence, increases the occupation of each vibrational mode. The density of
the condensate in the lowest virbational mode increases accordingly. Further, with Rabi coupling $\Omega \sim U$, we go beyond the
collisional blockade limit so that an arbitrary
number of bosons can occupy the AQD-site to make wide temporal and spatial fluctuation.
  
It is known that in the strong coupling regime the local spectrum can contain
  a bound or/and antibound one-particle states in addition to the
  continuum.\cite{logan} In this paper we select the coupling strength
  $\Omega$ such that these extra states do not occur, which is the only
  restriction for the coupling-strength $\Omega$.
  
As a final remark we note that the B-SIAM Hamiltonian conserves the total number of bosons. This is in contrast to the spin-boson model,\cite{Bulla2003,Bulla2005} where the bath contains excited phonons, the number of which is not conserved.

\section{The Bosonic NRG}
\label{sec:Method}\subsection{Mapping onto semi-infinite chain}
In this section we describe the numerical renormalization group (NRG) method
for conserved bosons, which is used to solve the B-SIAM Eq.~(\ref{bsiam-H})
introduced in the previous section. Details of NRG for bosons are presented in
the Appendices. This method is an adoption of the NRG from Ref.~(\onlinecite{Bulla2005}) to deal with bosons with a conserved number of particles.

As in the other NRG approaches,\cite{Bulla2008} the frequency range $[0,\omega_c]$ of the bosonic bath spectral function $J(\omega)$ is divided into intervals $[\omega_c\Lambda^{-(n+1)},\omega_c\Lambda^{-n}]$, where $\ n=0,1,2,...,$ and $\Lambda> 1$ is an NRG discretization parameter. The limit $\Lambda \rightarrow 1$ corresponds to the exact case.  Within each of these intervals the spectral function $J(\omega_c\Lambda^{-(n+1)}<\omega<\omega_c\Lambda^{-n})$ is approximated by its mean value 
\begin{eqnarray}
\bar{J}_n\equiv\frac{\int_{\omega_c\Lambda^{-(n+1)}}^{\omega_c\Lambda^{-n}}J(\omega)d\omega}{ (\omega_c\Lambda^{-n}-\omega_c\Lambda^{-(n+1)})}.
\end{eqnarray}
Next, following the same steps as in the spin-boson model in Refs.~(\onlinecite{Bulla2005,Lee_PhD}), we obtain a discretized version of the Hamiltonian (\ref{bsiam-H}) with new $V_n$ and $\epsilon_n$, which are defined on a discrete frequency grid, and with new bath bosonic operators labeled by discrete quantum numbers $n$. 
Now, this discretized model is mapped onto a semi-infinite chain~\cite{Bulla2003,Bulla2005,Bulla2008}  and we obtain the following  Hamiltonian
   \begin{eqnarray}
     H &=& \varepsilon b^\dagger b+\frac{U}{2}b^\dagger b (b^\dagger
     b-1)+V(b^\dagger \bar{b}_{0}+\bar{b}^\dagger_{0}b) \nonumber\\
     &+&\sum_{m=0}^{\infty}\varepsilon_m \bar{b}^\dagger_{m}\bar{b}_{m}+\sum_{m=0}^{\infty}t_m (\bar{b}^\dagger_{m}\bar{b}_{m+1}+\bar{b}^\dagger_{m+1}\bar{b}_{m}).
     \label{chain-H}
\end{eqnarray}
The bath degrees of freedom are represented by a tight-binding Hamiltonian with new creation and annihilation operators $\bar{b}_m^{(\dagger)}$ and $\bar{b}_m$, and the on-site energies $\varepsilon_m$, and hopping matrix elements $t_m$ between nearest neighbour sites. Both of them fall off exponentially, i.e. $t_m,\;\varepsilon_m \propto \Lambda^{-m}$.\cite{Bulla2003,Bulla2005,Bulla2008} Only the first site of the semi-infinite chain, which is denoted by the index $m=0$, is coupled to the impurity by the hybridization $V$. 

The Hamiltonian (\ref{chain-H}) cannot be diagonalized numerically for the
semi-infinite chain. Therefore, we need to truncate it at $m=M-2$, which
corresponds to taking $M$ sites, including the impurity-site, in the chain. Since the Hamiltonian parameters $t_m,\;\varepsilon_m$ decay exponentially with $m$, this truncation is justified at large $M$. The Hamiltonian diagonalized numerically has the form 
\begin{eqnarray}
  H_M &=& \varepsilon b^\dagger b+\frac{U}{2}b^\dagger b (b^\dagger
  b-1)+V(b^\dagger \bar{b}_{0}+\bar{b}^\dagger_{0}b) 
  \label{chain-H-M}
\\
\nonumber
 &+&\sum_{m=0}^{M-2}\varepsilon_m
  \bar{b}^\dagger_{m}\bar{b}_{m}+\sum_{m=0}^{M-3}t_m
  (\bar{b}^\dagger_{m}\bar{b}_{m+1}+\bar{b}^\dagger_{m+1}\bar{b}_{m}).
\end{eqnarray}
The Hamiltonian (\ref{chain-H-M}) commutes with the number operator
   \begin{eqnarray}
      N_{M}=b^\dagger b+\sum_{m=0}^{M-2}\bar{b}_m^\dagger\bar{b}_m.
     \label{qn}
\end{eqnarray}
Hence, the eigenstates of $H_M$ are also the eigenstates of $N_M$, so they are labeled by the corresponding quantum number $N$. The Hilbert space of all states with the same $N$ is denoted by $\mathcal{H}_N$. 
The dimension of each Hilbert space $\mathcal{H}_N$ with a given $M$ is 
    \begin{eqnarray}
     \mathcal{D}_{N} = \frac{(M-1+N)!}{(M-1)!\ N!}.
     \label{dimq}
\end{eqnarray}
%Here the total number of particles $N$ is not limited by the number of sites
%$M$, whereas the value in the fermionic system is $2M$ at most.
Unfortunately, for large $N$ and $M$ the Hilbert space dimension is so large
that direct diagonalization methods are not efficient. Therefore, the Hamiltonian (\ref{chain-H-M}) is diagonalized iteratively as is discussed next. 

\subsection{Iterative Diagonalization}

\label{itdiag}

At the beginning for small $M$ and $N$ such that the Hilbert space dimension $\mathcal{D}_{N}$ is less than typically few thousands, which depends on the computing facility, we perform exact diagonalization of the Hamiltonian (\ref{chain-H-M}) for a given M and all possible $N$ such that 
\begin{equation}
N=0,1,2,...,N_{\rm max},
\label{cutoff}
\end{equation}
where $N_{\rm max}$ is a cutoff for a number of particles. The truncation of the possible particle numbers, which is an approximation, is a necessary to make a computation feasible. As we will see later if the cutoff $N_{\rm max}$ is large enough then it does not affect obtained results. 

Having diagonalized the Hamiltonian $H_M$ for a given $M$ we increase the system size by adding one more site to the chain. Then we diagonalize the  Hamiltonian $H_{M+1}$ which has the form
\begin{eqnarray}
  H_{M+1} = H_M+ 
\varepsilon_{M-1} \bar{b}^\dagger_{M-1}\bar{b}_{M-1}+ \\ \nonumber 
 t_{M-2} (\bar{b}^\dagger_{M-2}\bar{b}_{M-1}+ \bar{b}^\dagger_{M-1}\bar{b}_{M-2}).
\end{eqnarray}
If  it turns out that the dimension of the Hilbert space is too large now, we need to construct an effective representation of the low energy eigenstates while $M$ increases. This is done iteratively as is described below.

We keep the dimension of the Hilbert space constant by taking only the low
energy eigenstates. However, to be able to make a direct comparison of the spectra while $M$ increases we need to scale the $M+1$ site Hamiltonian as follows
\begin{eqnarray}
  H_{M+1} = \Lambda H_M+ \Lambda^{M-1}\left[
\varepsilon_{M-1} \bar{b}^\dagger_{M-1}\bar{b}_{M-1}+\right. \\ \nonumber 
\left. t_{M-2} (\bar{b}^\dagger_{M-2}\bar{b}_{M-1}+ \bar{b}^\dagger_{M-1}\bar{b}_{M-2})
  \label{chain-H-M+1} \right],
\end{eqnarray}
where we keep the same symbol for the Hamiltonian. 
All eigenvalues of $H_M$ for all $ 0 \leq N\leq N_{\rm max}$ are sorted in an ascending way, and the $N_s$ 
eigenstates $|N,r_N\rangle_M$ with the lowest eigenvalues are used in diagonalizing $H_{M+1}$. Explicitly, we take into account such states that 
\begin{eqnarray}
  H_M|N,r_N\rangle_M&=&E_{r_N,M}(N)|N,r_N\rangle_M,
\label{diag_elements-N}
\end{eqnarray}
with $r_N=1,...,n_s^{(N)}$, 
where $n_s^{(N)}$ is the number of $N$-particle states with the lowest
eigenvalues $E_{r_N,M}(N)$ in each Hilbert space
$\mathcal{H}_N$.\cite{remark_on_N} The dimension of the Hilbert space $N_s$ is
given by the summation of $n_s^{(N)}$,
\begin{equation}
N_s=\sum_{N=0}^{N_{max}} n_s^{(N)},
\end{equation}
and optimized to perform the computation feasible.\cite{remark_on_cost}

In the Hilbert space of the $H_{M+1}$  Hamiltonian, the $N$-particle states are given by
\begin{eqnarray}
\{|N, R \rangle_{M+1} \} = \{ | N-k, r_{N-k}\rangle_M \otimes |k\rangle \}_{k=0,...,N},
\label{states}
\end{eqnarray}
where 
\begin{eqnarray}
|k\rangle = \frac{(\bar{b}_{M-1}^{\dagger})^{k}}{\sqrt{k!}}|0\rangle,
\end{eqnarray}
is the $k$-particle state on the $M-1$ site in the chain, and $|0\rangle$ is
an empty (vacuum) state on this last site. In Eq.~(\ref{states}) the
quantum number $R\equiv r_{N-k}$ means the quantum number of the $H_M$
Hamiltonian with $N-k$ particles. The numbers $R$ are not the quantum numbers
labeling the eigenstates of the Hamiltonian $H_{M+1}$. This is due to the fact
that the new Hamiltonian $H_{M+1}$ does not commute with the total number of particles $N_M$ of the previous system with the Hamiltonian $H_M$, i.e. we can check that 
$[H_{M+1}, N_M]\neq 0$, where $N_M$ is defined in (\ref{qn}).
In order to find eigenstates of $H_{M+1}$ in a basis (\ref{states}) we construct the Hamiltonian matrix elements
\begin{equation}
H(R;R^\prime)\equiv _{M+1}\langle N,R|H_{M+1}|N,R^\prime\rangle_{M+1},
\label{newelements}
\end{equation} 
and diagonalize this matrix obtaining a set of eigenvalues and eigenstates
\begin{equation}
|N,\omega_N\rangle_{M+1}=\sum_{R} U_N(\omega_N;R)|N,R\rangle_{M+1}
\label{egbasis}
\end{equation}
where $U_N(\omega_N;R)$ is an orthogonal matrix, and $\omega_N$ are new quantum numbers labeling an $N$ particle eigenstate of $H_{M+1}$ with eigenvalue $E_{\omega_N,M+1}(N)$.
The procedure described from Eqs.~\eqref{states} to \eqref{egbasis} is repeated for all  $N=0,1,2,...,N_{max}$.

In the next iteration step we extend the system by adding one more site to the
chain and use the eigenstates (\ref{egbasis}) of $H_{M+1}$ to construct a
basis of the new Hamiltonian in a way analogous to (\ref{states}). Repeating
the same procedure as described between Eqs.~\eqref{diag_elements-N} to \eqref{egbasis} we obtain new eigenstates and eigenvalues of the larger system. Further details on the iterative diagonalization is presented in Appendix~\ref{app-bsiam:QNiteration}.

We proceed iterative diagonalizations until the many particle spectra
approach the trivial fixed point of the non-interacting bosonic
bath. The low-energy spectrum of Mott and BEC phases and the structure of the
fixed points are presented in Sec.~\ref{subsec:gd}.

%Those issues are crucial not only for the implementation of the NRG but also
%for the intepretation of the numerical results. Indeed, they present an
%important criterion to distinguish one phase from the other of the phase
%diagram in Fig.~\ref{bsiam_fig1}. In the next section, we discuss the issues
%focussing on the
%ground states of the Hamailtonian $H_M$ in equation~(\ref{chain-H-M}).

\section{\label{sec:phases} Zero-temperature phase-diagram}
\subsection{Overview}
\begin{figure}
\begin{center}
  \includegraphics[clip=true,width=0.4\textwidth]{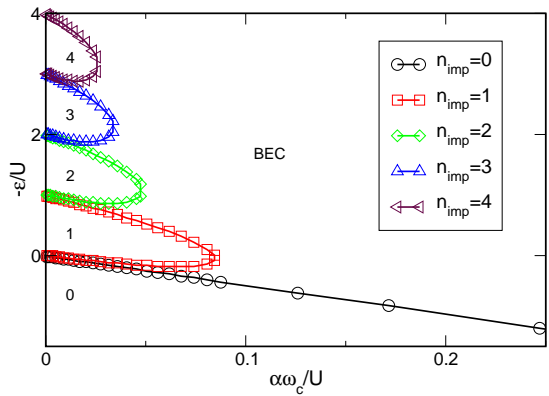}
\end{center}
\caption{Zero-temperature phase diagram of the B-SIAM for bath exponent
  $s=0.4$ and fixed impurity Coulomb interaction $U=0.5\omega_c$. The
  different symbols denote the phase boundaries between Mott phases and the
  BEC phase. The Mott phases are labeled by the number of the impurity-quasiparticle, $n_{imp}$. Only the Mott phases with $n_{imp}\leq 4$ are shown. The NRG parameters are $\Lambda=2.0$, $N_b=10$, and $N_s=100$.}
\label{bsiam_fig1}
\end{figure}
 \begin{figure}
\begin{center}
\vspace{0.cm}
  \includegraphics[angle=0,width=0.47\textwidth]{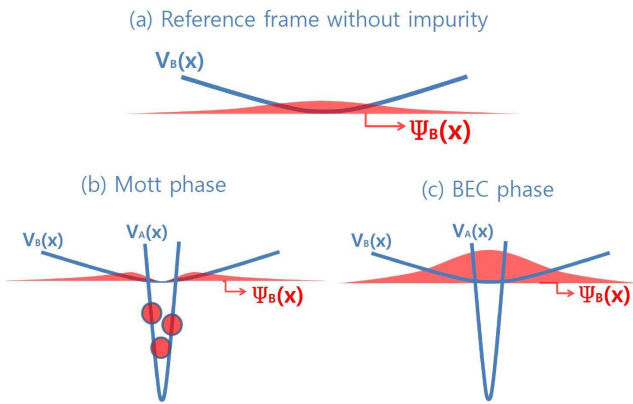}
\vspace{-0.cm}
\end{center}
\caption{(a) Reference frame: A non-interacting BEC state, $\Psi_B({\bf x})$,
  is confined in a shallow trapping potential $V_B({\bf x})$. \\
(b) Mott phase:
  The impurity-quasiparticle consists of an integer number of depleted particles, (depicted as balls), which are tightly trapped in $V_A({\bf
  x})$. The other bosons contained in $V_B({\bf x})$ still form a BEC cloud but the local density of the condensate vanishes in the
vicinity of the AQD. (c) BEC phase: The impurity-quasiparticle forms
 a part of a BEC state to enhance the density of the condensate around the
 AQD. The confining potential $V_A({\bf x})$ and $V_B({\bf x})$ are in all three
  directions with spherical symmetry.}
\label{opttrap2}
\end{figure}
The zero-temperature phase diagram in Fig.~\ref{bsiam_fig1} is calculated for fixed $U=0.5\omega_c$ with
the parameter space spanned by the dimensionless coupling constant
$\alpha=\frac{(1+s)}{2}\Omega^2$ and the impurity energy $\varepsilon$. 
We choose $s=0.4$ as the exponent of the power law in $J(\omega)$ in
Eq.~(\ref{bath-spctdst1}). A similar phase diagram for different bath exponent
$s=0.6$ has been presented in
Ref.~(\onlinecite{Lee2007}). The phase diagram is characterized by a sequence of lobes. We use the
terminology ``{\it Mott phase}'' for the inside of the lobes and ``{\it BEC phase}'' for the region outside of the
lobes. 

The Mott and the BEC phases are distinguished by a hybridized state
that is formed around the AQD as illustrated in Fig.~\ref{opttrap2}. Fig.~\ref{opttrap2}-(a) shows a BEC state of a
non-interacting bosonic bath, where all existing particles occupy the lowest vibrational mode of the
shallow potential $V_B({\bf x})$.  In the presence of the AQD, however, particles around the AQD can be either completely depleted
(Fig.~\ref{opttrap2}-(b)) or even more concentrated toward the local site
(Fig.~\ref{opttrap2}-(c)).  We call the collective excitation around the AQD
as {\it impurity-quasiparticle}.

In the Mott phase, the impurity-quasiparticle consist of an integer number of
depleted particles, (depicted as balls in Fig.~\ref{opttrap2}-(b)), which are tightly trapped in $V_A({\bf
  x})$. The number of the depleted particles is used to label the different Mott phases in the phase diagram in
Fig.~\ref{bsiam_fig1}. The other bosons contained in $V_B({\bf x})$ still form a BEC cloud but the local density of the condensate vanishes in the
vicinity of the AQD.

 In the BEC phase (Fig.~\ref{opttrap2}-(c)), the impurity-quasiparticle forms
 a part of a BEC state to enhance the density of the condensate around the
 AQD. The enhancement of the condensate-density is due to the strong Raman
 coupling $\Omega$ and the deep attractive potential $\varepsilon<0$ of the AQD. 

Numerical evidences for our assertions are presented in
the rest part of the paper. In Sec.~\ref{subsec:gd}, we look into the
contribution of the impurity-quasiparticle to the ground state energy. In Sec.~\ref{sec:localdynamics}, the
calculation of the local Greens function is presented to show the local
dynamics of normal and condensate particles.

\subsection{Impurity contribution to the ground state energy}
\label{subsec:gd}
The ground state energy of a non-interacting bosonic bath is zero since all
existing particles occupy the lowest vibrational mode with zero-energy. An impurity site with repulsive interaction $U$, however, can deplete some
particles from the zero-energy mode and shift the ground state energy to be finite. In general, the non-zero ground state energy depends on the
total number of particles ($N$) in the system.
\begin{figure}
\begin{center}
  \includegraphics[clip=true,width=0.36\textwidth]{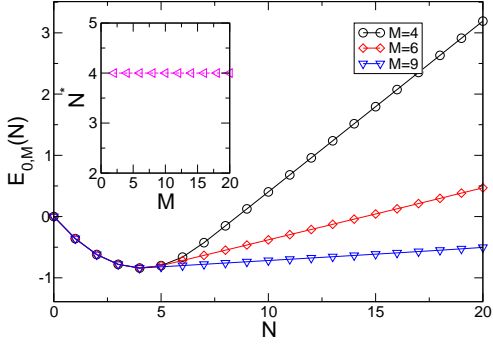}
\end{center}
\caption{Ground state energy $E_{0,M}(N)$ vs. $N$ calculated for $s=0.6,
  U=0.1,\varepsilon=-0.36$ and $V=0.01$. The parameters are
  chosen inside of the Mott lobe labeled by 4 (Mott phase 4). The inset shows the
  position of the minimum point $N=N^*$ as a function of $M$. 
The NRG parameters used are $\Lambda=1.5$ and $N_s=1000$.}
\label{gd-mott}
\end{figure}
\begin{figure}
\begin{center}
  \includegraphics[clip=true,width=0.36\textwidth]{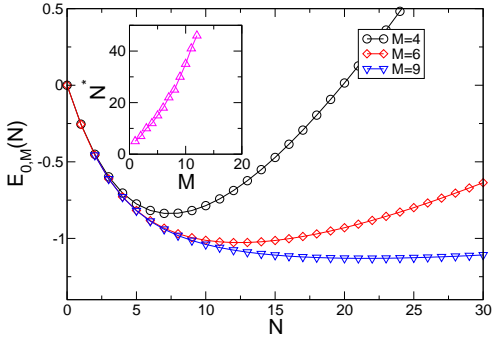}
\end{center}
\caption{Ground state energy $E_{0,M}(N)$ vs. $N$ calculated for $s=0.6,
  U=0.1,\varepsilon=-0.05$ and $V=0.4$. The parameters are chosen
  outside of the Mott lobes (BEC phase). The inset shows the
  position of the minimum point $N=N^*$ as a function of $M$. The NRG parameters used are $\Lambda=1.5$ and $N_s=1000$.}
\label{gd-bec}
\end{figure}
\begin{figure}
\begin{center}
  \includegraphics[clip=true,width=0.36\textwidth]{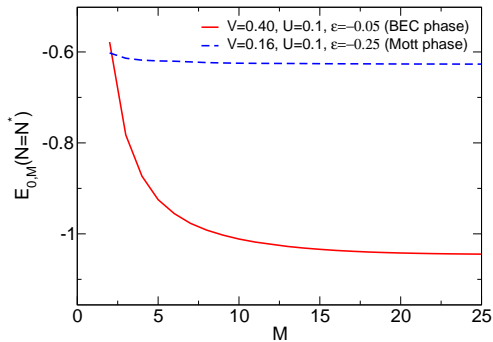}
\end{center}
\caption{The minimum of the ground state energy $E_{0,M}(N)$ at $N=N^*$ as a function of
  $M$. The solid and dashed lines correspond to the BEC and the Mott phases,
  respectively. The bath exponent is fixed to $s=0.6$. The NRG
  parameters used are $\Lambda=1.25$, $N_{max}=40 \ (8)$ and $N_s=8000 \ (600)$ for a BEC (Mott) phase.}
\label{gdvsM}
\end{figure}

 Fig.~\ref{gd-mott} and Fig.~\ref{gd-bec} show the $N$-dependence of the
 ground state energy $E_{0,M}(N)$ in Mott and BEC phases, respectively. The different curves are the results from different size ($M$)
of the systems.
 The ground state energy $E_{0,M}(N)$
decreases until the configuration around the AQD,
(i.e. impurity-quasiparticle) is optimized. The occupation at the minimum
point is denoted by $N^*$.

The minimum ground state energy $E_{0,M}(N)$ at $N=N^*$ is plotted as a
function of the system size $M$ in Fig.~\ref{gdvsM}. The minimum ground state
energy $E_{0,M}(N)$ at $N=N^*$ converges in the limit $M\rightarrow \infty$. 
%The convergence of
%$E_{0,M}(N^*)$ is easily attained in a Mott phase
%whereas, in a BEC phase, the ground state energy converges after a large
%number of iterations, which one can also observe in Fig.~\ref{gd-mott} and
%Fig.~\ref{gd-bec} with comparing three curves for different $M$.
Once the system converges into the large $M$ limit, all ground states for different
$N$ become degenerate. Indeed, Fig.~\ref{gd-mott} and Fig.~\ref{gd-bec} show that the ground state
energy $E_{0,M}(N)$ becomes almost independent of $N$ already for $M=9$. 

In the thermodynamic limit ($N\rightarrow \infty$, $M\rightarrow \infty$), the
result of adding (or removing) one particle is to convert a
state of a system of $N$ particles into the same state of a system of $N\pm 1$
particles:
\begin{equation}
 \lim_{N\rightarrow \infty}\lim_{M\rightarrow \infty}|N\pm1,0\rangle_{M}=
 \lim_{N\rightarrow \infty}\lim_{M\rightarrow \infty}|N,0\rangle_M.
 \label{deg}
\end{equation}
Here $|N,0\rangle_M$ is the $N$-particle ground states of $H_M$. 
This is the case of a condensate consisting of a macroscopic
number of particles, i.e. a coherent state.~\cite{Lifshitz1980}

The degenerate feature of the ground states in Eq.~(\ref{deg}) is extended to the low lying excited states when the
many-particle spectrum reaches a fixed point.

\begin{figure}
\begin{center}
\includegraphics[clip=true,width=0.5\textwidth]{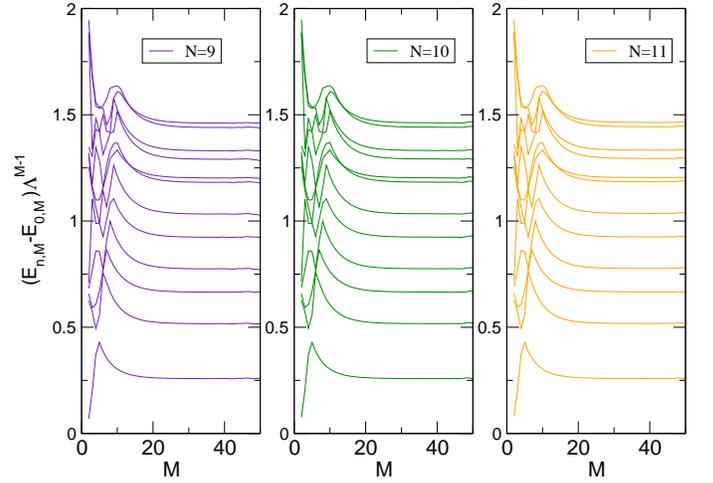}
\end{center}
\caption{The lowest lying many-particle levels $E_{n,M}\Lambda^{M-1}$ versus
  iteration number $M$ for parameters $s=0.7$, $V=0.01$, $U=0.5$, and
  $\varepsilon=-1.2$ (Mott phase). The NRG parameters used are $\Lambda=1.25$, $N_s=3000$
  and $N_{max}=15$.}
\label{fdQNmott}
\end{figure}
\begin{figure}
\begin{center}
\includegraphics[clip=true,width=0.5\textwidth]{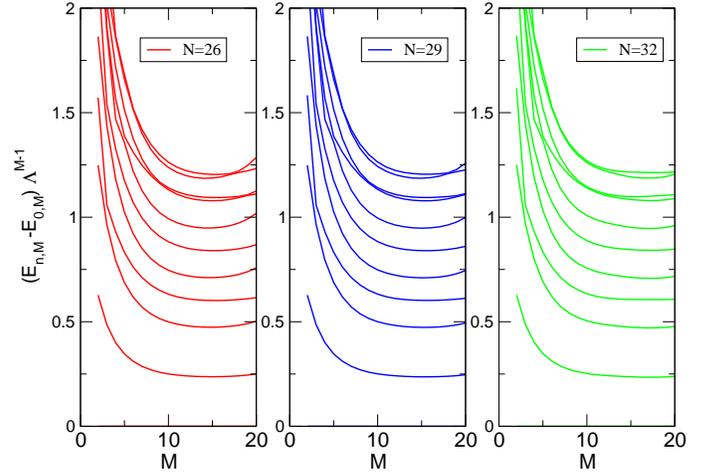}
\end{center}
\caption{ Flow diagram of the lowest lying many-particle levels $E_{n,M}\Lambda^{M-1}$ versus
  iteration number $M$ for parameters $s=0.7$, $V=0.4$, $U=0.1$ and
  $\varepsilon=-0.05$. The NRG parameters used are $\Lambda=1.25$, $N_s=8000$
  and $N_{max}=40$.}
\label{fdQNbec}
\end{figure}
\begin{figure}
\begin{center}
\includegraphics[clip=true,width=0.38\textwidth]{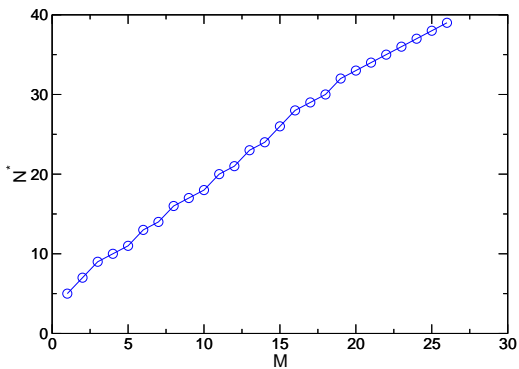}
\end{center}
\caption{The quasiparticle-occupation $N^*$ versus the system-size $M$. The
  data obtained for parameter parameters $s=0.7$, $V=0.4$, $U=0.1$, and
  $\varepsilon=-0.05$ (BEC phase). The NRG parameters are $\Lambda=1.25$,
  $N_{max}=40$ and $N_{s}=8000$.}
\label{nstar}
\end{figure}

 Fig.~\ref{fdQNmott} shows the energy flow of the lowest lying many-particle
levels $E_{n,M}(N)$ versus iteration number $M$. The parameters $V,U$ and
$\varepsilon$ are chosen for the system to flow into a Mott phase.
Three pannels show the $N$-particle eigenstates for $N=9,10,$ and $11$. The
eigenstates in the three figures flow into the same fixed point, which is a
trivial fixed point of a non-interacting bosonic bath. It means that the dynamics of the AQD, i.e.
the impurity-quasiparticle, is suppressed in this energy-scale so that the
low-lying excitations show the dynamics of the non-interacting bosons that locate far from the AQD-site.

 Fig.~\ref{fdQNbec} shows the lowest lying many-particle levels in a BEC phase. 
Three pannels show the $N$-particle eigenstates for $N=26,29,$ and $N=32$, which
flow into the same strong-coupling fixed point. The level-spacing in the
strong-coupling fixed point is different from the one in the non-interacting
fixed point - the reason is not clear yet.

As a last remark, we mention the conditions for numerical convergence. We see that the energy-levels start
to deviate from the strong-coupling fixed point around at the iterative step $M=20$. The upturn
(deviation from the fixed point) appears if the number of particles $N$ is not
large enough compared to $N^*(M)$. The $N^*$ increases with increasing $M$
(see Fig.~\ref{nstar}) and reaches the value $N^*\sim 30$ at the iteration
$M=20$.  The $N$-particle eigenstates flows into the same strong-coupling fixed
point only if $N$ is larger than $N^*$.

 The quick and the slow convergence in Mott and BEC phases respectively can be interpreted as
 following. The system-size $M$ corresponds to the number of vibrational
 modes that are taken into account in $H_M$, i.e. the larger system involves
 more vibrational modes with small energy. From this we can conclude that the
 impurity-quasiparticle in a Mott phase consists of the depleted particles
 occupying the higher vibrational modes in $V_B({\bf x})$, which can be
 described by a relatively small system size. In a BEC phase, however, the
 impurity-quasiparticle is a part of a condensate which consists of a
 macroscopic number of particles with almost zero-energy. Thus one needs a
 large value of $N$ and $M$ to properly describe the condensate.
 
\section{Local dynamics at zero temperature}
\label{sec:localdynamics}
The local Green's function of the impurity model is defined
as
\begin{eqnarray}
G(z)&=&\frac{1}{i}\int_0^\infty dt e^{izt}\langle[b(t),b^\dagger]\rangle.
\label{greenftn}
\end{eqnarray}
where $b^{(\dagger)}$ is an annihilation (creation) operator for the impurity. 
 \begin{figure}
\begin{center}
  \includegraphics[clip=true,width=0.45\textwidth]{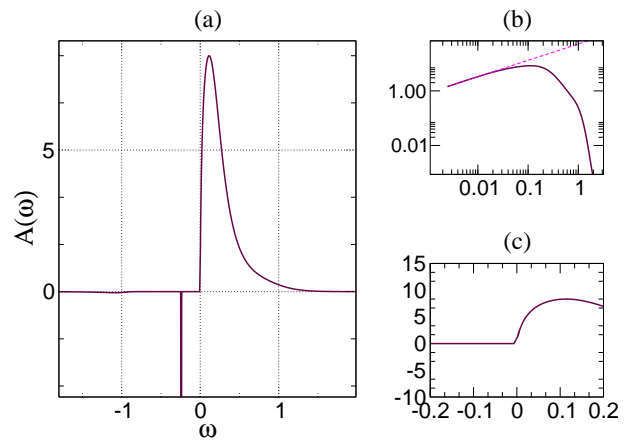}
\end{center}
\caption{(a) The local spectral density of the B-SIAM for bath exponent
  $s=0.6$ and fixed impurity Coulomb interaction $U=0.5\omega_c$, the onsite
  impurity energy $\varepsilon=-0.7$, and the hybridization $V=0.15$ (Mott
  phase 2).  The NRG parameters are $\Lambda=1.25$, $N_{max}=3$, and
  $N_s=1000$.
 (b) The (positive) low-frequency part of $A(\omega)$
  (solid line)
  in log-log scale. The dashed line is a guide line for eyes showing a
  power-law behavior ($\propto \omega^{s}$, $s=0.6$). (c) The low-frequency
  part of $A(\omega)$ in linear scale. $A(\omega)$ vanishes at $\omega=0$. }
\label{imgmott}
\end{figure}
\begin{figure}
\begin{center}
  \includegraphics[clip=true,width=0.45\textwidth]{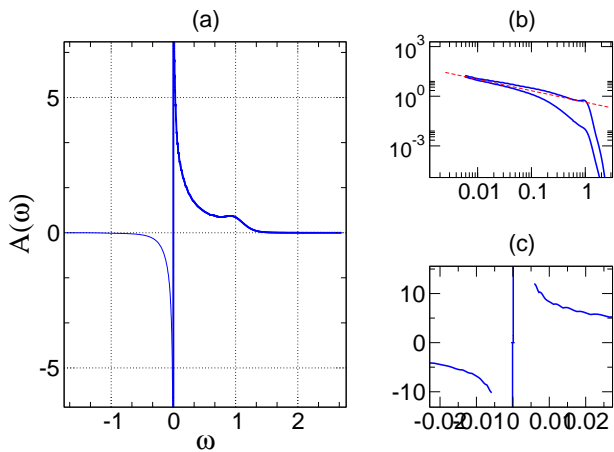}
\end{center}
\caption{(a) The local spectral density of the B-SIAM for bath exponent
  $s=0.6$ and fixed impurity Coulomb interaction $U=0.1\omega_c$, the onsite
  impurity energy $\varepsilon=-0.05$, and $V=0.3$ (BEC phase). The NRG
  parameters are $\Lambda=1.25$, $N_{max}=40$, and $N_s=5000$. (b) The low-frequency part of $|A(\omega)|$ as a function of $|\omega|$ in log-log scale, where upper and lower curves correspond to the positive and
  negative spectral density in the first and third quadrant, respectively. The
  dashed line in the inset is a guide line for eyes showing a
  power-law behavior ($\propto \omega^{-s}$, $s=0.6$). (c) The low-frequency
  part of $A(\omega)$ in linear scale. $A(\omega)$ shows two $\delta$-peaks at
  $\omega\approx \pm 0.0001$. The position of two peaks approach $\omega=0$ in
thermodynamic limit $M\rightarrow\infty$ as seen in Fig.~\ref{img-mu}.}
\label{imgbec}
\end{figure}
The local spectral density $A(\omega)$ is the imaginary part of the
local Green's function,
\begin{equation}
A(\omega)=-\frac{1}{\pi}\Im G(\omega+i\delta).
\label{localspct}
\end{equation} 

The local spectral density in a Mott phase (Fig.~\ref{imgmott}) shows two
quasiparticle peaks that are separated by a gap, $\Delta_{gap} \sim
0.2$.  A sharp peak at $\omega\sim -0.2$ is a signal of hole-excitation in the
AQD. The particles trapped in the AQD show no resonance with the reservoir as if they are isolated from it. In fact,
most of particles in the reservoir are immobile since they are condensed at
zero-energy and make no resonance with the particles in
the AQD.

 The local occupation at the AQD-site can be obtained by integrating the spectral weight
below the chemical potential $\mu=0$,
\begin{eqnarray}
%n_{loc}&\equiv&\langle\Psi_{0,M}|\hat{b}^\dagger\hat{b}|\Psi_{0,M}\rangle=0.999\nonumber
n_{loc}(T=0)&=&\left[\int_{-\infty}^{\infty} f_{BE}(\omega)A(\omega)
  d\omega\right]_{T=0}\nonumber\\
 &=&1.8875
\label{localoccup1}
\end{eqnarray}
where the Bose-Einstein distribution function $f_{BE}(\omega)$ is given as a
step function at zero temperature,
\begin{eqnarray}
\lim_{\beta\rightarrow \infty}f_{BE}(\omega)=\lim_{\beta\rightarrow \infty}
\frac{1}{e^{\beta \omega}-1}=-\Theta(-\omega)\nonumber \\
\label{bedtb}
\end{eqnarray}
with $\Theta(\omega)=1 $ for $\omega>0$ and $\Theta(\omega)=0 $ for $\omega<0$ .

\begin{figure}
\begin{center}
  \includegraphics[clip=true,width=0.45\textwidth]{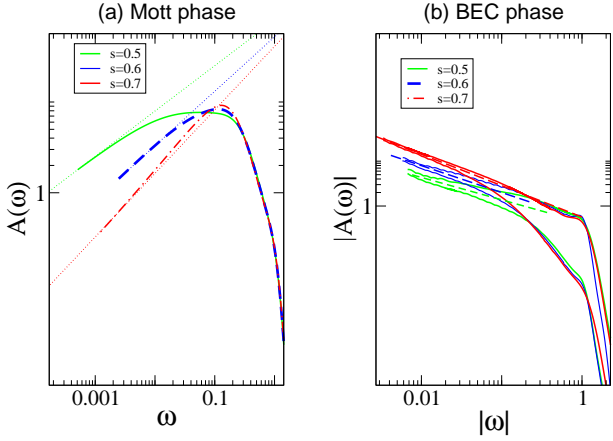}
\end{center}
\caption{(a) The low-frequency behavior of
  $A(\omega)$ for $U=0.5\omega_c$, $\varepsilon=-0.7$, $V=0.15$ (Mott phase 2) and for various
  bath-exponent  $s=0.5,\ 0.6,$ and $0.7$. (b) The low-frequency behavior of $A(\omega)$ for $U=0.1\omega_c$, $\varepsilon=-0.05$, $V=0.3$ (BEC phase) and for various  bath-exponent $s=0.5,\ 0.6,$ and $0.7$.  The NRG parameters are
  $\Lambda=1.25$, $N_{max}=40$, and $N_s=5000$. The  dashed lines are guide lines
  for eyes to show the power-law behavior. The NRG parameters are $\Lambda=1.25$, $N_{max}=3$, and $N_s=1000$.
}
\label{imgbec-sdep}
\end{figure}

 Creating a
particle at the AQD-site gives a broad peak at positive frequency. There is no
feature at $\omega=0$ (Fig.~\ref{imgmott}-(c)) indicating that the BEC is locally forbidden around the AQD-site. The
$A(\omega)$ vanishes at $\omega=0$ with a power-law behavior,
\begin{equation}
A(\omega) \propto \omega^{s},  \ \omega>0,
\label{lowfreq1}
\end{equation}
 which is the same for the bath spectral function $J(\omega)$ in
 Eq.~(\ref{bath-spctdst1}). The power-law behavior for various bath exponents $s$ is shown in Fig.~\ref{imgbec-sdep}-(a).

Fig.~\ref{imgbec} shows the spectral density in the BEC phase. The spectral density in the BEC phase (Fig.~\ref{imgbec}) diverges at $\omega=0$,
\begin{equation}
A(\omega) \propto {\rm sgn}(\omega)|\omega|^{-s},
\label{lf-spctdst}
\end{equation}
where the power-law corresponds to the inverse of the bath
spectral density $J(\omega)$, see Fig.~\ref{imgbec}-(b). The divergence of
$A(\omega)$ occurs if a hybridized state is pinned at the gapless point of the
spectral function $J(\omega)$. 
To discuss more details, let us look into the local Green's function $G(z)$ written as
\begin{equation}
  G(z)=(z-\varepsilon-\Sigma(z))^{-1}
\label{gimp-sfu1}
\end{equation}
where $\varepsilon$ is the energy of the impurity level (with operator
$b^{(\dagger)}$) and  $\Sigma(z)$ is the total self-energy of the
impurity model. The imaginary part of the Green's function in Eq.~(\ref{gimp-sfu1}) is given as
\begin{equation}
\Im\left[
  G(z)\right]=\frac{\Im\left[\Sigma(z)\right]}{(\Re\left[z-\varepsilon-\Sigma(z)\right])^2+(\Im\left[\Sigma(z)\right])^2}
\label{form_img1}
\end{equation}
 with  $z=\omega+i0^+$. The actual calculation of $\Sigma(z)$ is in progress and will be presented in our
subsequent paper. Here we assume that the imaginary part of the self-energy $\Im\left[\Sigma(z)\right]$ follows a power-law
behavior with the same exponent as the bath spectral function
$J(\omega)\propto \omega^s$:
\begin{equation}
\Im\left[\Sigma(z)\right] \propto \omega^s.
\label{gapless-sfe}
\end{equation}
The singular behavior of the local spectral density
 $A(\omega)$ in Eq.~(\ref{lf-spctdst}) can appear when the impurity bound state occurs at $\omega=0$:
 \begin{equation}
\Re\left[\omega-\varepsilon-\Sigma(\omega+i0^+)\right]=0 \ {\rm at \ } \omega=0.
\label{omega_bs}
\end{equation}
 The imaginary part of the self-energy shows a power-law behavior as assumed
 in Eq.~\eqref{gapless-sfe}.
In the case, the $\Im\left[G(z)\right]$ becomes inverse-proportional to $\Im\left[\Sigma(z)\right]$,
\begin{equation}
\Im\left[G(z)\right] \propto \frac{1}{\Im\left[\Sigma(z)\right]} \propto \omega^{-s}.
\end{equation}
 If the impurity bound state occurs below the chemical potential, the first
 term in the numerator in Eq.~(\ref{form_img1}) is non-zero at $\omega=0$,
 which makes $\Im\left[G(z)\right]$ proportional to
 $\Im\left[\Sigma(z)\right]$ around the gapless point $\omega=0$,
\begin{equation}
\Im\left[G(z)\right] \propto \Im\left[\Sigma(z)\right] \propto \omega^{s}.
\end{equation}
A similar feature of $A(\omega)$ is observed in the pseudo-gap
Anderson model,~\cite{remark_on_sgAm,Bulla2000} where a Kondo bound state appears at the
gapless Fermi level.

 The singular behavior of $A(\omega)$ for various bath
exponents $s$ is shown in Fig.~\ref{imgbec-sdep}-(b).
 
\begin{figure}
\begin{center}
  \includegraphics[clip=true,width=0.45\textwidth]{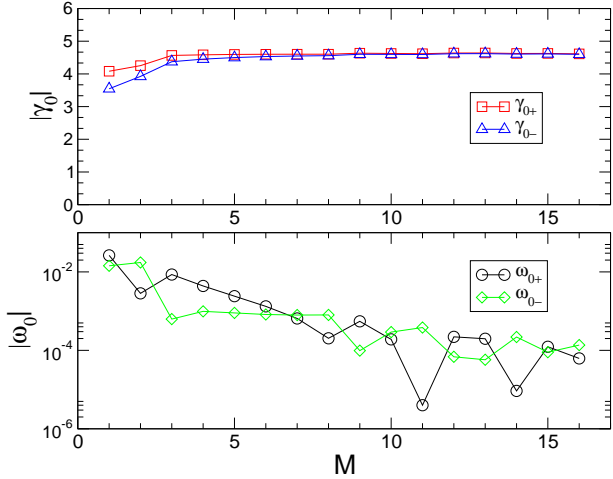}
\end{center}
\caption{The position ($|\omega_{0\pm}|$) and the amplitude ($|\gamma_{0\pm}|$) of
  the two peaks in Fig.~\ref{imgbec}-(c) depending on the size of system
  $M$. The indices $0+$ ($0-$) denote the peak at positive (negative) frequency, respectively. 
}
\label{img-mu}
\end{figure}

Another interesting feature in the BEC phase is the finite spectral weight at
 $\omega=0$ as shown in Fig.~\ref{imgbec}-(c).  Fig.~\ref{imgbec}-(c) shows two peaks at small frequency
 $\omega_{0\pm}\approx\pm 0.0001$ with opposite sign of spectral weight. In the limit $M\rightarrow \infty$, the position of both peaks approaches to zero
 ($\omega=0$) and the amplitude $|\gamma_{0\pm}|$ converges
 to the same value (Fig.~\ref{img-mu}). The finite spectral weight at
 $\omega=0$ indicates the existence of the condensate particles in the AQD-site.

The local occupation at the AQD-site can be obtained from integrating the spectral weight
below the chemical potential $\mu=0$,
\begin{eqnarray}
%n_{loc}&\equiv&\langle\Psi_{0,M}|\hat{b}^\dagger\hat{b}|\Psi_{0,M}\rangle=0.999\nonumber
n_{loc}(T=0)&=&\left[\int_{-\infty}^{\infty} f_{BE}(\omega)A(\omega)
  d\omega\right]_{T=0}\nonumber\\
&=&\gamma_{0-}+ \lim_{\epsilon\rightarrow 0-}\int_{-\infty}^{\epsilon} A(\omega)
  d\omega\nonumber\\
 &=&4.6+0.22
\label{localoccup-bec}
\end{eqnarray}
where the Bose-Einstein distribution function $f_{BE}(\omega)$ at zero
temperature is given in equation~(\ref{bedtb}). The first term ($=4.6$) in
equation~\eqref{localoccup-bec} is the contribution of the condensate
particles whereas the second term($=0.22$) is the contribution of
particles that are depleted from the condensate. 

\section{Conclusion}
\label{sec:conclusion}
 The bosonic single-impurity Anderson model is studied to understand the local
dynamics of an atomic quantum dot (AQD) coupled to a BEC state. The major
result presented in this paper is the calculation of the impurity Green
function but, in addition, considerable space is devoted to refine the description of the Mott and
the BEC phases. The local collisional interaction, dominant over the Raman
coupling, depletes the particles around the
AQD out of the condensate (Mott phase). Otherwise, the Raman transition makes
the density of the BEC state even more concentrated toward the local site (BEC
phase). The AQD can share a coherent phase of the macroscopic condensate only
in the BEC phase and can be used to probe the
decoherence of the BEC state.~\cite{Ng2008,Recati2005,Bruderer2006} 

The scheme for the quantum dense coding protocol,~\cite{Heaney2009A} requires two separate AQDs, both of which are coupled to the same BEC
state. In Ref.~(\onlinecite{Heaney2009A}), it is assumed that a signal between the two
AQDs is phase-locked through a BEC state with uniform density and
phase. However the phase preserved in each AQD can depend on the
position of the dots when the AQDs make the BEC state
non-uniform. In this case, the spatial fluctuation of a BEC cloud in the presence of
two AQDs deserves of further research, for which a recent extension of the NRG
technique, computing spatial correlation function for the Kondo
screening cloud,~\cite{Borda2007} is also applicable.

\section{Acknowledgment}
We have benefited from discussions with Gun-Sang Jeon, Ki-Seok Kim, Tetsuya
Takimoto, Dieter Vollhardt, Xin Wan and Philipp Werner. Special
thanks to Vincent Sacksteder for his help on program optimization. This research was supported by the DFG through SFB 484, SFB 608, FOR 960, and TRR 80. H.-J. Lee acknowledges the Max Planck Society and Korea Ministry of Education, Science and Technology for the joint support of
the Independent Junior Research Group at the Asia Pacific Center for
Theoretical Physics. KB acknowledges the grant N N202 103138 of Polish Ministry of Science and Education.

\appendix
\section{Details about the Iterative Diagonalization}
\label{app-bsiam:QNiteration}
Now we obtain the matrix elements in Eq.~(\ref{newelements}),
\begin{equation}
H(R;R^\prime)\equiv _{M+1}\langle N,R|H_{M+1}|N,R^\prime\rangle_{M+1},
\end{equation} 
where the $N$-particle states $|N,R\rangle_{M+1}$ is defined in Eq.~(\ref{states}).

It is straightforward to demonstrate that the diagonal matrix elements of
$H_{M+1}$ are
\begin{eqnarray}
H(R;R)=E_{R,M}(N-k) + \ k\ \varepsilon_{M-1}.
\label{elements_H_NI-N2}
\end{eqnarray}

The only non-vanishing off-diagonal elements of $H_{M+1}$ are given by
\begin{eqnarray}
&&H(R^\prime;R)\nonumber\\
&&= \delta_{k^\prime,k-1} t_{M-2}\sqrt{k} \ _{M}\langle N-k^\prime,R^\prime || \bar{b}_{M-2}^\dagger
|| N-k,R\rangle_M \ \nonumber \\
&&+\delta_{k^\prime,k+1} t_{M-2}\sqrt{k+1} \ _M\langle N-k^\prime,R^\prime || \bar{b}_{M-2} || N-k,R\rangle_M,\nonumber \\
&&
\label{offdiag_elements-N2}
\end{eqnarray}
where $\langle || \bar{b}_{M-2}^{(\dagger)} || \rangle$ are the invariant matrix elements.

In obtaining Eq.~(\ref{elements_H_NI-N2}), we have made use of the
following results,
\begin{eqnarray}
&&_{M+1}\langle N^\prime,R^\prime || \bar{b}_{M-1} ||
N,R\rangle_{M+1}=\delta_{k^\prime, k-1}\sqrt{k}
\label{bm-1}
\end{eqnarray}
and
\begin{eqnarray}
_{M+1}\langle N^\prime,R^\prime || \bar{b}_{M-1}^\dagger || N,R\rangle_{M+1}=\delta_{k^\prime, k+1}\sqrt{k+1}.
\end{eqnarray}
 which follow from the definition of the basis set in Eq.~(\ref{states}).

From Eq.~(\ref{elements_H_NI-N2}) and Eq.~(\ref{offdiag_elements-N2}), it is clear that we can set up the matrix of $H(R;R^\prime)$ starting with the
knowledge of the previous iterative step such as the eigenenergy $E_{R,M}(N-k)$ and the matrix elements
\begin{equation}
_{M}\langle N-k-1,r^\prime ||\bar{b}_{M-2}||N-k,r\rangle_M
\end{equation}
 for $k=0,...,N$.

The actual iteration upon entering the stage $(M+1)$ would proceed as follows. We first start with the lowest allowed value of $N_{M+1}(=0)$, and then increase it in steps of $1$. Within a given $\mathcal{K}_N$ subspace, we construct the matrix
\begin{equation}
H(R;R^\prime)\equiv _{M+1}\langle N,R|H_{M+1}|N,R^\prime\rangle_{M+1}.
\end{equation} 
Diagonalization of this matrix gives a set of eigenstates
\begin{equation}
|N,\omega_N\rangle_{M+1}=\sum_{R} U_N(\omega_N;R)|N,R\rangle_{M+1}
\end{equation}
where $U_N$ will be an orthogonal matrix. The diagonalization means no more than the knowledge of $E_{R,M+1}(N)$ and $U_N(\omega_N;R)$. After completing the diagonalization for one $N$, we proceed up, increasing $N$ in steps of $1$.
In order to go to the next iteration we need to calculate $_{M+1}\langle N-1,\omega^{\prime} || \bar{b}_{M-1} ||N,\omega \rangle_{M+1}$.
 Using the results in Eq.~(\ref{bm-1}), it is easy to verify that
\begin{eqnarray}
&&_{M+1}\langle N-1,\omega_{N-1}^{\prime} || \bar{b}_{M-1} ||N,\omega_N \rangle_{M+1}\nonumber \\
&=&\sum_{R} U_{N-1}(\omega_{N-1}^\prime;R)U_{N}(\omega_N;R)\sqrt{k}
\end{eqnarray} 
where $k$ is the number of particles on the $M-1$ site in the chain as given
in the Eq.~(\ref{states}).

\section{Calculation of local spectral density}
The NRG method uses a discretized version of the Anderson model in a
semi-infinite chain form  in Eq.~(\ref{chain-H-M}). The resulting
spectral functions will therefore be given as a set of discrete
$\delta$-peaks. For example, the spectral representations of the one-particle Green's function $G(z)$ is
\begin{eqnarray}
A(\omega)&=&-\frac{1}{\pi} \Im G(\omega) \nonumber\\
&&=\sum_{N,r}\sum_{N^\prime,r^\prime}|\langle N,r ||b^\dagger||N^\prime,r^\prime\rangle|^2\exp{\{-\beta E(N,r)\}}\nonumber \\
&&\ \ \ \ \ \ \ \ \ \ \ \ \ \ \ \times\delta(\omega-E(N,r)+E(N^\prime,r^\prime))\nonumber \\
&&-\sum_{N,r}\sum_{N^\prime,r^\prime}|\langle N,r ||b||N^\prime,r^\prime\rangle|^2\exp{\{-\beta E(N,r)\}}\nonumber\\
&&\ \ \ \ \ \ \ \ \ \ \ \ \ \ \ \times\delta(\omega+E(N,r)-E(N^\prime,r^\prime))\nonumber\\
\label{gcn-green2}
\end{eqnarray}
Here $|N,r\rangle$ and $E(N,r)$ are the abbreviation of $|N,r\rangle_M$ and
$E_M(N,r)$ in Eq.~(\ref{diag_elements-N}).

As a practical matter, however, calculating the states of $H_M$ for large $N$
is hard to deal with because the number of $N$-particle states of $H_M$
explodes in combinatorial way as shown in Eq.~(\ref{dimq}). Thus we
introduce cut-off, 
\begin{eqnarray}
\sum_{N}&\rightarrow& \sum_{N=0}^{N_{max}}.
\label{sum-cutoff1}
\end{eqnarray}
The value of $N_{max}$ has to be larger than the minimum point of
the ground state energy at $N=N^*$.

At zero temperature, the ensemble average in Eq.~(\ref{gcn-green2}) is replaced to the ground expectation value $\langle N^*,0 |...|N^*,0\rangle$ :
\begin{eqnarray}
A(\omega)_{T=0}&=&-\frac{1}{\pi} \Im G(\omega)_{T=0} \nonumber\\
&=&\sum_{r}|\langle N^*+1,r|b^\dagger|N^*,0\rangle|^2\nonumber\\
&&\times\delta(\omega-E(N^*+1,r)+E(N^*,0))\nonumber \\
&&- \sum_{r}|\langle N^*-1,r|b|N^*,0\rangle|^2\nonumber \\
&&\times\delta(\omega+E(N^*-1,r)-E(N^*,0)). \nonumber \\
\label{cn-green-zeroT1}
\end{eqnarray}
 The matrix elements $\langle N,r ||b^\dagger||N^\prime,r^\prime\rangle$ and the energies $E(N,r)$ are calculated in the NRG method.
The resulting spectral function, as a set of $\delta$-functions at frequencies $\omega_n$ with weights $g_n$, are broadened on a logarithmic scale as
   \begin{equation}
     g_n     \delta(\omega-\omega_n) \rightarrow g_n\frac{e^{-b_n^2/4}}{b_n\omega_n\sqrt{\pi}}\exp[-\frac{(\ln\omega-\ln \omega_n)^2}{b_n^2}].
    \end{equation}

 In our calculations, the width $b_n$ is chosen as b independent of $n$ and
 the typical values we use are in the range $0.01<b<0.1$. A $\delta$-peak in
 Fig.~\ref{imgmott}-(a) is an intrinsic $\delta$-peak without any resonance,
 for which we use a value, $b_n=0.0001$.


\begin{thebibliography}{99}
\bibitem{remark_on_N} The set of the quantum numbers $r_N$ is different for different $N$. However, they are usualy represented by integers which formally should have a subscript $N$. To avoid this cumbersome notation we add explicitly $N$ in the eigenvalues $E_{r_N,M}(N)$ and later for particular cases we use simple integers for $r_N$.
\end{thebibliography}
\end{document}